\documentclass[preprint]{aastex}

\usepackage{epsf}
\usepackage{amsmath}
\usepackage{amssymb}
\usepackage{graphics}
\usepackage{graphicx}
\usepackage{cite}

\def\gsim{\mathrel{\raise0.35ex\hbox{$\scriptstyle >$}\kern-0.6em
\lower0.40ex\hbox{{$\scriptstyle \sim$}}}}
\def\lsim{\mathrel{\raise0.35ex\hbox{$\scriptstyle <$}\kern-0.6em
\lower0.40ex\hbox{{$\scriptstyle \sim$}}}}

\begin{document}

\title{Microwave Emission from the Edgeworth-Kuiper Belt and the Asteroid Belt 
Constrained from WMAP}

\author{Kazuhide Ichikawa$^{a}$\footnote{Present address: Department of Micro Engineering,
Kyoto University, Kyoto 606-8501, Japan} and Masataka Fukugita$^{a,b,c}$}
\affil{$^{a}$Institute for Cosmic Ray Research, University of Tokyo,
           Kashiwa 277-8582, Japan\\
$^{b}$Institute for Advanced Study, Princeton, NJ 08540 USA\\
$^{c}$Institute for the Physics and Mathematics of the Universe,
University of Tokyo,
           Kashiwa 277-8583, Japan\\}

\begin{abstract}

Objects in the Edgeworth-Kuiper belt and the main asteroid belt should
emit microwaves that may give rise to extra anisotropy signals in the
multipole of the cosmic microwave background (CMB) experiment.
Constraints are derived from the absence of positive detection of such
anisotropies for $\ell\lesssim 50$, giving the total mass of
Edgeworth-Kuiper belt objects to be smaller than 0.2$M_\oplus$. This
limit is consistent with the mass extrapolated from the observable
population with the size of $a\gtrsim 15$ km, assuming that the
small-object population follows the power law in size $dN/da\sim
a^{-q}$ with the canonical index expected for collisional equilibrium,
$q\simeq 3.5$, with which 23\% of the mass is ascribed to objects
smaller than are observationally accessible down to grains.  A similar
argument applied to the main asteroid belt indicates that the grain
population should not increase faster than $q\simeq 3.6$ towards
smaller radii, if it follows the power law continued to observed
asteroids with larger radii. It is underlined that both cases are at
or only slightly above the limit that can be physically significant,
implying the importance of tightening further the CMB anisotropy
limit, which may be attained with the observation at higher radio
frequencies.

\end{abstract}
%\keywords{cosmology}

%%%%%%%%%%%%%%%%%%%%%
\section{Introduction} \label{sec:introduction}
%%%%%%%%%%%%%%%%%%%%%
Whether exists a substantial population of small bodies in the
Edgeworth-Kuiper belt is an interesting question to ask, whichever is
the origin, either remaining from the original planetary nebula or
produced from the interaction among planetesimals or asteroids.  The
direct search for objects in the Edgeworth-Kuiper belt (we refer to them
in brevity as KBO) reaches to 30 km in diameter. It is observed
that KBO with the size larger than $a_{\rm br}\approx 40-100$\,km
decreases towards a larger size as $a^{-q}$ with $q\approx 4-5$
\citep{Trujillo2001,Bernstein2004,Fuentes2008, Fraser2009} (We denote by $a$
the effective radius).  The size
distribution shows a break towards smaller radii at around $a_{\rm
 br}\approx 40-100$km, and then flattens to be $q\approx 3$.  With the occultation
giving the Fresnel diffraction one may observe objects smaller than a km
\citep{Bailey1976, Dyson1992}; for recent observational efforts, see
 \citet{Bianco2010}. Using such a technique a small KBO is recently
discovered at around the radius of 300 m \citep{Schlichting2009},
which points towards a low-mass slope $q\approx 3.9\pm0.3$. These slopes and
the presence of the break are understood from the consideration that
they are caused by frequent destructive collisions for smaller bodies
\citep{KenyonBrom2004,PanSari2005}.

\citet{PanSari2005} further argue that the slope becomes steeper again at
a smaller radius and it becomes the Dohnanyi power $q\sim 3.5$
\citep{Dohnanyi1969} of collisional equilibrium, when the bodies held
together matter-strength dominated at around $a\lsim 100$ m rather
than gravitationally dominated at larger radii.  
This is 
in fact the slope preferred
for submicron size dust grains to account for the extinction law in optical
wavelengths (Mathis, Rumpl \& Nordsieck 1977; Weingartner \& Draine 2001).
It is interesting to see that this power of the size distribution is  
consistent with micron-size grains
observed by the Ulysses and Galileo satellites at the Jupiter distance
\citep{Frisch1999}, while the agreement with the power of the size of 
interstellar dust may be merely accidental .  
It is an interesting question if this
distribution continues to super-micron sizes \citep{Draine2009}.

Small size grains may have fallen to the Sun by the Poynting-Robertson
drag, and even smaller grains may have been swept away by solar winds.
The action of the Poynting-Robertson drag implies the mm size grains
as the minimum size remaining in the Edgeworth-Kuiper belt
after the age of the Solar System ellapsed.  The direct search by the
Ulysses and Galileo satellites, however, shows the presence of 
$\mu$m-size grains even in the inner heliosphere around the Jupiter distance
\citep{Frisch1999}, whereas the Poynting-Robertson drag gives the
typical falling age to be only $5\times10^4$ yr for micron-size
grains.

Small bodies may aggregate to form larger bodies or may be eroded into
smaller bodies.  The issue of small bodies would hint us to understand
the dynamics of planet formation. A possible way to explore small
bodies is to look for the infrared emission from those objects
\citep{Backman1995, Teplitz1999, KenyonWindhorst2001}.  The difficulty
with the infrared emission to explore the Edgeworth-Kuiper belt
objects is that the emission is largely overcome by that from the
inter-planetary dust and the asteroid belt, and the subtraction of these
components is not easy.

The temperature of grains in the Edgeworth-Kuiper belt is 
about 40K, if we assume them to be blackbody, so that
they also emit microwaves in the
Rayleigh-Jeans region. Such emission might be detectable in the cosmic
microwave background experiment, the precision results being
already available and more to come in the near future.
In fact, a constraint from radio emission from such objects was recently
considered in \citet{Babich2007} for the distortion of the CMB spectrum
constrained by the FIRAS (Far-InfraRed Absolute Spectrometer) of the COBE
satellite \citep{Fixsen1996}.

In this paper, we consider what we can learn as to the distribution of
planetesimals and grains in the Edgeworth-Kuiper Belt, and also
in the main asteroid belt of the Solar System from the CMB anisotropy
data of Wilkinson Microwave Anisotropy Probe (WMAP)
\citep{Bennett:2003bz, Hinshaw2007}. 
 WMAP has measured
  the microwave emission in the sky, capable of making the full-sky
  temperature maps in five bands: 22.8, 33.0, 40.7, 60.8 and 93.5\,GHz
  (or correspondingly the wavelengths of 13, 9.1, 7.3, 4.9 and 3.2\,mm).
While this does not give spectral information, it probes at a high
accuracy the angular distribution, to which objects at the
Edgeworth-Kuiper belt distance (and the main asteroid belt) would
contribute.  We aim at deriving a constraint on the total mass of the
KBO and, in addition, the objects in the main asteroid belt 
including small grains as a
function of the assumed power index of the size distribution for small
objects to grains.  We consider the grain size larger than 1$\mu$m,
which is large enough so that atomic and molecular excitation is
unimportant and the black body approximation may apply to heating and
radiation.  We bear in mind the question what observational strategies
would be useful to explore deeper Edgeworth-Kuiper belt objects, and
also consider if future CMB experiments could detect the emission from
the Edgeworth-Kuiper Belt or the main asteroid belt.

%%%%%%%%%%%%%%%%%%%%%%%%%%%%%%%%%%%%%%%
\section{Size distribution and thermal radiation from grains}  \label{sec:KBO}
%%%%%%%%%%%%%%%%%%%%%%%%%%%%%%%%%%%%%%%

We consider the mass distribution $n(M)$ 
in the broken power law $n(M)=K M^{-\alpha}$ with two different
values of $\alpha$ for $M_{\rm min} < M < M_{\rm br}$ and
$M_{\rm br} < M < M_{\rm max}$ with a break at $M_{\rm br}$ for KBO,
as observationally indicated.
We take the density of the objects to be constant at 
$\rho = 2.5\,{\rm g~cm}^{-3}$.
The power index $\alpha$ is then related to that of
the size distribution $q$ in $dN/da\sim a^{-q}$ as $\alpha=(q+2)/3$, and 
to the slope of the brightness number count of the object
$dN/dm\sim 10^{b(m-m_0)}$ as $b=(q-1)/5$ if the albedo is
constant.  

We may take the broken power law with two breaks as
expected by a theoretical argument (e.g., Pan \& Sari 2005),
where the second break is suggested to take place at a 100 m size.
The observation for KBO shows at least one break at $\approx40-100$ km
radius, which is ascribed to frequent destructive collisions effective for
smaller KBO.
Smaller than the break size, however, the size distribution
is poorly determined;  the current result varies around
$q=2.8\pm0.6$ \citep{Bernstein2004,Fuentes2008} to $q\approx 3.9\pm 0.3$
(Schlichting 2009), which sandwich the \lq canonical' power
of collisional equilibrium  $q=3.5$ ($\alpha=11/6$).
In view of this present observational uncertainties we adopt for 
KBO a broken power law 
with a single break at a 100 km radius, rather than introducing the second
break at a smaller radius in order not to 
increase the number of unconstrained parameters.
(For the main asteroid belt, the size distribution below the first break
is reasonably determined, and we take a two-break power law; see later.)

For a larger size, we take $q=4.5$, i.e., $\alpha=13/6$, consistent
with the observation. For a smaller size we leave the faint end slope
as a free parameter, with the canonical collisional equilibrium value 
$q=3.5$ or $\alpha=11/6$ in mind. 
The upper cutoff of the integral is taken to be
the mass of Pluto, $M_{\rm max} = 0.0022 M_\oplus$
(1080 km in radius for our density $\rho=2.5$\,g cm$^{-3}$). 
Our results are insensitive to the 
maximum cutoff as we take $q=4.5$.  We take the break radius to be 100 km, or
correspondingly the mass $M_{\rm br} = 1\times 10^{22}$g, consistently with
\citet{Bernstein2004}.

We take two alternative choices for the minimum mass of the integral. One 
corresponds to $a_{\rm min}=1$\,mm and the other corresponds 
to $a_{\rm min}=1$\,$\mu$m. The former corresponds roughly to the mass
that receives the Poynting-Robertson drag for grains to fall to the Sun in 
the age of the Sun $4.7\times 10^9$ yr at the Edgeworth-Kuiper belt distance.
The other is the case where a significant amount of the grain mass is included.
Grains with the size smaller than 1$\mu$m are likely to be removed
by radiation pressure in a dynamical time.
We take the total mass as a parameter: 
\begin{eqnarray}
M_{\rm tot} &=& \int_{M_{\rm min}}^{M_{\rm max}} M n(M) dM.
\end{eqnarray}

%\begin{figure}[htbp]
\begin{figure}[!b]
\begin{center}
\includegraphics[scale = 0.8]{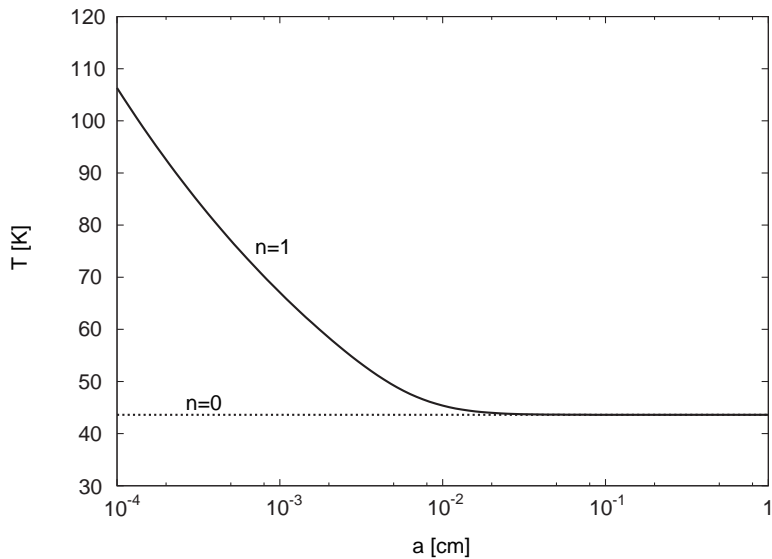}
\caption{The temperature of grains at the Edgeworth-Kuiper belt distance 
as a function of their radius. $n=1$ is assumed for 
the power of the suppression factor
in Eq. (\ref{eq.3}). 
We assume $A=0.04$ and $D = 40$\,AU.}
\label{fig:dust_temp}
\end{center}
\end{figure}

We take the Edgeworth-Kuiper belt distance at $D=40$ AU, since
the consideration of a more detailed distribution would simply leave
more unconstrained parameters. This simplification is sufficient for
us to consider the effect.  We calculate the thermal emission from
KBOs, assuming that it arises from the conversion of the solar
radiation absorbed by the object into IR and microwave emission,
ignoring any internal sources of energy.  The equation of balance for
the object with radius $a$, albedo $A$ and emissivity suppression
factor $\epsilon$ at the heliocentric distance $D$ is written
\begin{eqnarray}
\frac{L_\odot}{4\pi D^2} \cdot \pi a^2 \cdot (1-A) = \int_0^\infty \pi B_\nu(T) \epsilon(\nu,a) d\nu \cdot 4\pi a^2, \label{eq:balance}
\end{eqnarray}
where 
$B_\nu(T)$ stands for the flux of the black body radiation of frequency $\nu$
at temperature $T$, 
and $\epsilon$ is the suppression factor  
when the wavelength of radiation is larger than the size of the object, which 
is assumed to be
\begin{eqnarray}
\epsilon(\nu, a)= 
\begin{cases} 1 & \text{if $a>\lambda$,}
\\
(a/\lambda)^n &\text{if $a<\lambda$,}
\label{eq.3}
\end{cases}
\end{eqnarray}
where $\lambda$ is the wavelength.

We set $n=1$ as in the formal factor in Mie scattering
in the following; see e.g., \citet{Spitzer1978,Backman1995}.  
This gives the temperature of objects at the Edgeworth-Kuiper belt distance as
in Fig.~\ref{fig:dust_temp}, showing that for emission in WMAP frequencies
this suppression causes some effects 
only for $a<50\mu$m, the emission from which is unimportant
for the parameters that concern us, and
hence the results hardly depend on the suppression factor assumed
for a wide variety of its choice.  
We take $A=0.04$.  
For large size grains
$T_{\rm KBO}=43.7$\,K. 
The size of grains we consider is large enough, so that shot noise heating by
a single photon is unimportant, and hence we do not treat atomic
emission as was considered in detail in \citet{WeingDraine2001}.

\begin{figure}[htbp]
\begin{center}
\includegraphics[scale = 1.5]{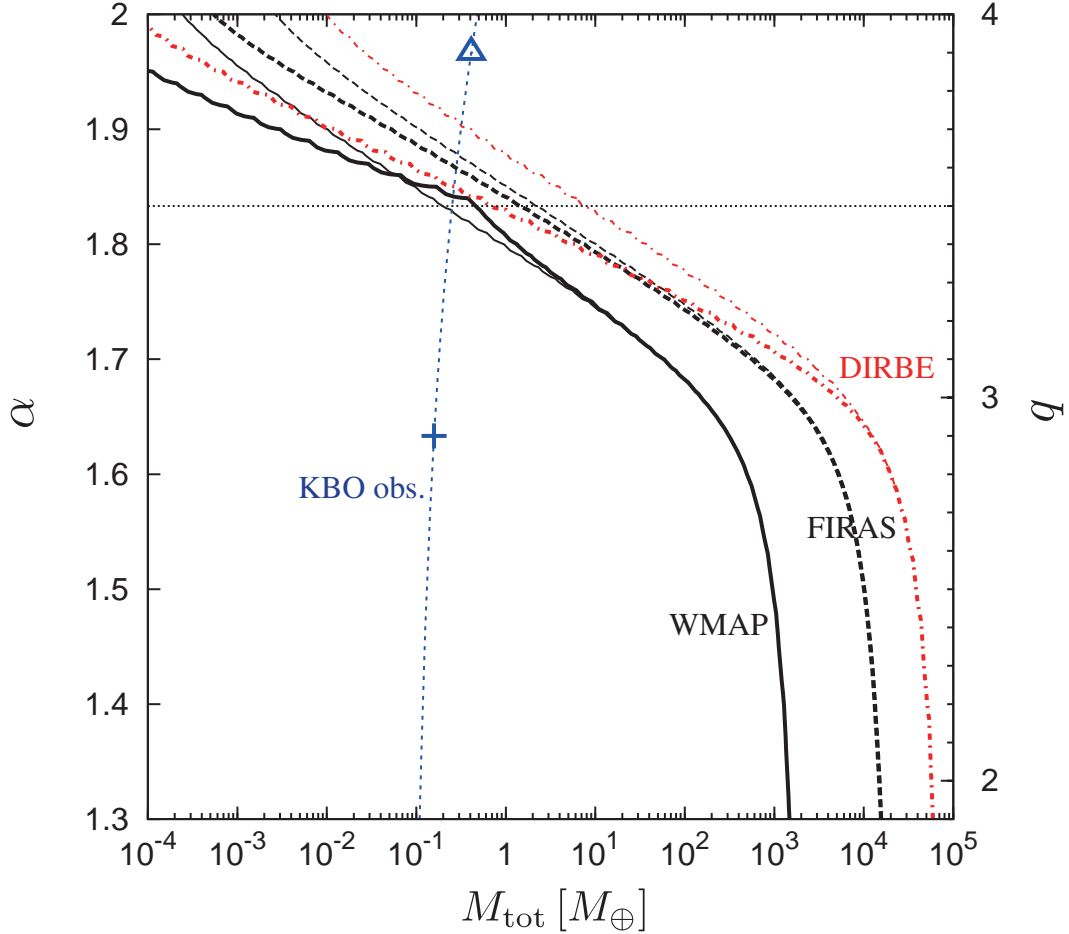}
\caption{Limits on the total mass in the Edgeworth Kuiper belt as a
 function of the small-mass slope $\alpha$ of the mass function at
 the 95\% C.L. CMB anisotropy constraints from WMAP are shown with
 solid curves and spectral constraints from COBE/FIRAS are shown with
 dashed curves.  The allowed regions are above the curves.  Thinner
 lines are for $a_{\rm min}=1$\,mm and thicker lines for $a_{\rm
   min}=1$\,$\mu$m.  Limits from the FIR brightness of COBE/DIRBE
 observations are added with dash-dotted curves. The nearly vertical
 dotted line is the total mass of the objects in the Edgeworth-Kuiper
 belt extrapolated from the direct observation of larger size objects
 with the assumed power index of $\alpha$, the plus symbol being the
 power index estimated from the observation of objects smaller than
 the break radius, $a\approx 100$ km \citep{Bernstein2004}.  The
 triangle on the curve indicates the central value of the power index
 inferred from the object found in the occultation observation
 \citep{Schlichting2009}.  The horizontal dotted line refers to the
 canonical power of collisional equilibrium $q=3.5$ of the size
 distribution.}
\label{fig:summary}
\end{center}
\end{figure}

%%%%%%%%%%%%%%%%%%%%%%%%%%%%%
\section{Contribution to CMB anisotropies}  \label{sec:COBE}
%%%%%%%%%%%%%%%%%%%%%%%%%%%%%
Let us first consider the contribution of the KBO to spectral data
averaged over the sky, such as those obtained by COBE/FIRAS. This has
already been done in \citet{Babich2007}.  We consider the two
cases for $a_{\rm min}=1$\,mm, as assumed in Babich et al., and for
$a_{\rm min}=1$\,$\mu$m.

The KBO emission would modify the CMB spectrum averaged over the sky
%\begin{eqnarray}
$4\pi D^2 B_\nu(T_{\rm CMB})$ as 
$(4\pi D^2 -A_{\rm tot}) 
B_\nu(T_{\rm CMB}) + A_{\rm tot} B_\nu (T_{\rm KBO})$, 
where the effect is taken into account that KBOs block CMB photons over 
the area $A_{\rm tot}$ that is covered by KBOs, 
\begin{eqnarray}
A_{\rm tot} = \int_{a_{\rm min}}^{a_{\rm max}} \pi a^2 n(a) da 
%= \int_{M_{\rm min}}^{M_{\rm max}} \pi R(M)^2 n_m(M) dM.
\end{eqnarray}
The spectral distortion is written 
\begin{eqnarray}
\Delta B_\nu &=& \frac{1}{4\pi D^2} \left\{ \int_{a_{\rm min}}^{a_{\rm max}} 
\pi a^2 n(a) B_\nu (T(a)) \epsilon(\nu,a) da - A_{\rm tot} B_\nu(T_{\rm CMB}) 
\right\}. \label{eq:deltaBnu_dust}
\end{eqnarray}

Following \citet{Fixsen1996}, we calculate $\chi^2$ by minimizing over
the shift in the CMB temperature $\Delta T$ and the normalization
factor  $G_0$ of the Galactic emission contamination templet
prepared by Fixsen et al. 
To constrain the KBO contribution,
we allow $\Delta T$ and $G_0$
as free parameters to compute 
$\chi^2 = (\Delta B_\nu +\Delta T \frac{\partial B_\nu}{\partial T} + G_0 g(\nu))^2/\sigma_\nu^2$, 
where $g(\nu)$ is the modeled Galaxy emission spectrum: see \citet{Fixsen1996}
for details. 
 We show the limit
at the 95\% C.L.  ($\Delta \chi^2 = 5.99$) in Fig.~\ref{fig:summary}.
The two curves, labelled with FIRAS, shown with dashed correspond to
$a_{\rm min}=1\mu$m (thick curve) and 1mm (thin curve).  The
constraint with $a_{\rm min}=1$mm, of course, agrees with that of
\citet{Babich2007}, when the parameters used here are converted to
theirs.  The extension of the minimum cutoff radius from 1mm to
1$\mu$m modifies the result only little, no more than by a factor of 2
in mass units, unless $\alpha$ is close to 2.

We now turn to the calculation of the CMB anisotropy. We evaluate
\begin{eqnarray}
\Delta B_\nu ({\bf n}) = \frac{f({\bf n})}{D^2} \left\{ \int_{a_{\rm min}}^{a_{\rm max}} \pi a^2 n(a) B_\nu (T(a)) \epsilon(\nu,a) da - A_{\rm tot} B_\nu(T_{\rm CMB}) \right\}, \label{eq:deltaBnu_dust_WMAP}
\end{eqnarray}
where $f({\bf n})$ is the spatial distribution of
KBOs in the direction of ${\bf n}$, normalized to yield unity when
integrated over ${\bf n}$: $f({\bf n}) A_{\rm tot}\, d\Omega$
gives the area covered by KBOs at the direction ${\bf n}$.
We then convert the KBO contribution to brightness 
$\Delta B_\nu ({\bf n})$ to the temperature variation
$\Delta T({\bf n})$ using the black body radiation formula. We then 
write it in the form $\Delta T({\bf n}) = c_\nu f({\bf n})$. 
Writing the KBO distribution in terms of the spherical harmonics,
$f({\bf n})= \sum_{\ell,m} f_{\ell m} Y_{\ell m}({\bf n})$, we
represent
the coefficient of the harmonic decomposition of $\Delta T({\bf n})$ 
as  $a^{\rm KBO}_{\ell m} = c_\nu f_{\ell m} $.

Following \citet{Brown2001} we take the spatial distribution of 
KBOs to be double Gaussian around 
the ecliptic plane 
\begin{eqnarray}
f({\bf n}) \propto \sum_i A_i \exp \left\{ 
-\frac{(\theta - \theta_{\rm ecl})^2}{2 \sigma_{{\rm KBO},i}^2} \right\},
\end{eqnarray}
where
$\theta_{\rm ecl} \sim 60^\circ$
is the inclination angle between the ecliptic plane 
and the Galactic plane. 
We take the widths of the distribution of clasical KBO's, 
$\sigma_{{\rm KBO},1} = 2^\circ.2$ 
and $\sigma_{{\rm KBO},2} = 17^\circ$
with the ratio of the coefficients $A_1:A_2=0.65:0.35$
from the observed latitudinal distribution.
%\footnote{\citet{Brown2001} reported that the inclination distribution is 
%described by a sum of two Gaussian functions with the widths $2.6^\circ$ 
%and $15^\circ$. We adopt effective single Gaussian for the latitudinal 
%distribution for simplicity, which is sufficient for us.}.

\begin{figure}[!t]
\begin{center}
\includegraphics[scale = 1]{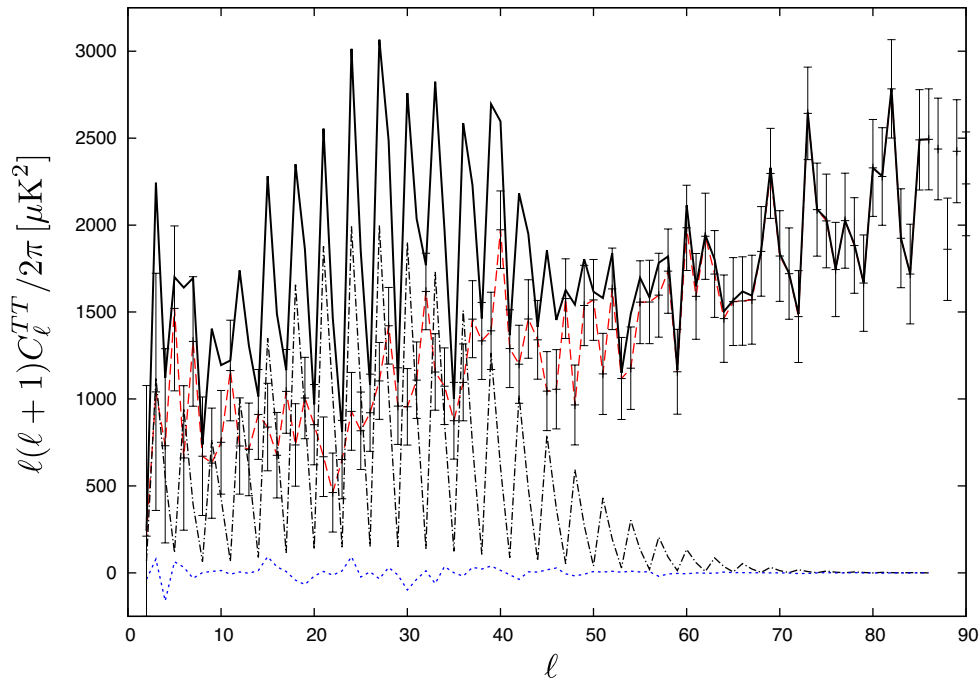}
\caption{Example of multipoles of temperature anisotropy. 
The red dashed line shows $C_\ell^{\rm CMB}$, the dot-dashed line shows
$C_\ell^{\rm KBO}$ and blue dotted line indicates the cross 
term  $C_\ell^{\rm cross}$, where the
parameters for KBO are taken to be 
$M_{\rm tot} = 1 M_\oplus$ and $\alpha = 1.83$, the
parameters allowed by a COBE/FIRAS spectral analysis but
is forbidden by WMAP anistropies. 
The total multipole is shown by the solid line $C_\ell$. 
The error bars depict the cosmic variance.}
\label{fig:cl}
\end{center}
\end{figure}

The multipole coefficient of the CMB power spectrum is 
modified by emission from KBOs as
\begin{eqnarray}
C_{\ell} &=& \frac{1}{2\ell +1} \sum_{m = -\ell}^{\ell} | a_{\ell m}^{\rm CMB} 
+ a_{\ell m}^{\rm KBO}|^2, \nonumber \\
&=& C_\ell^{\rm CMB} + C_\ell^{\rm KBO} + \frac{1}{2\ell +1} 
\sum_{m = -\ell}^{\ell} 2 {\rm Re}\, a_{\ell m}^{\rm CMB} 
a_{\ell m}^{\ast \rm \, KBO}. 
\label{eq:Cl}
\end{eqnarray}
We show an example of harmonics $C_\ell$ at $\nu= 93.5$\,GHz, the
highest frequency band of WMAP. The contribution expected from KBO
emission is shown for the parameters,  $M_{\rm tot} = 1 M_\oplus$ and $\alpha = 1.83$ in
Fig.~\ref{fig:cl}, which represent a case marginally allowed by COBE/FIRAS,
but will be rejected by the WMAP observation as shown in what follows 
(see Fig.~\ref{fig:summary}).  The KBO contribution shows the maximal
power at around, $\ell \approx 20-30$.  
The emission from the KBO for $\ell\lesssim 100$, with the choice 
of the trial parameters taken here, is only slightly larger than
the cosmic variance. 
The cross correlation signal, $C_\ell^{\rm cross}$, given by 
the last term in Eq.(\ref{eq:Cl}),
is much smaller.

With the lack
of any signals indicating the contribution of KBO emission we may set
$a_{\ell m}^{\rm CMB}$ to the observed value.
The error bars attached to the CMB data are the cosmic variance
\begin{eqnarray}
\frac{\sigma_\ell}{C_\ell^{\rm CMB}} = \sqrt{\frac{2}{2 \ell +1}}.
\end{eqnarray}
The accuracy of the WMAP observation for $\ell\lesssim 100$ that concerns
us here already reaches the cosmic variance.

We may derive the constraint from the CMB data, by requiring 
that the KBO contributions and the interference term between
CMB and KBO,
$C_\ell^{\rm KBO} + C_\ell^{\rm cross}$ 
be smaller than the cosmic variance. The result does not change if we
replace the cosmic variance with the observed WMAP data.
We calculate $\chi^2$ as
\begin{eqnarray}
\chi^2 = \sum_\nu \sum_\ell \frac{(C_\ell^{\rm KBO}+C_\ell^{\rm cross})^2}{\sigma_\ell^2},
\end{eqnarray}
where the summation $\nu$ is over the data in the three highest
frequency bands $\nu= 93.5$\,GHz, 60.8\,GHz and 40.7\,GHz of the WMAP.
The result is shown with solid curves in Fig.~\ref{fig:summary} above
for both $a_{\rm min}=1\,\mu$m (thick curve) and 1\,mm (thin curve).
The limit is derived basically from the highest frequency data, and
the inclusion of lower frequency data modifies little the result.  The
curve for $a_{\rm min}=1\,\mu$m generally appears lower in the figure
(i.e., the limit being tighter) than that for $a_{\rm min}=1$mm, but
the two curves are reversed at a specific $\alpha$, where the
integrand of Eq.~\eqref{eq:deltaBnu_dust_WMAP} vanishes for some value
of $\alpha$ owing to the presence of the suppression factor: $\Delta
B_\nu$ in Eq. (\ref{eq:deltaBnu_dust}) can be negative at some value
of $\nu$ for small grains ($a<<1$mm), which in turn allows extra
emission from larger objects given the total emission as the
constraint.  For the canonical power index $q=3.5$, the limit of the
KBO with $a_{\rm min}=1$mm is 0.2\,$M_\oplus$, of which 23 \% of the
mass arises from the objects below the limit of the observed KBO $a
\approx 15$ km. The KBO radiation arises dominantly from objects with
$a\leq 15$ km if $q\geq 3.2$.  The anisotropy data give a limit
stronger than the spectral distortion by a factor of $\approx 5$ in
the total mass.

In the same figure, we also display the limit from the FIR emission
using the COBE/DIRBE data which are already discussed by
\citet{Teplitz1999}. We take the observed FIR data \citep{Hauser1998}
to derive the bound.  The limit on the KBO emission should actually be
tighter, if the FIR emission from interplanetary dust and the asteroid
belt which give rise to the FIR signal.could properly be subtracted.
The reliable subtraction to give a small component, such as cosmic
infrared background, however, is notoriously difficult.  So, we take
here the observed brightness without subtraction.  If we would take
the modest subtraction, for example, of \citet{Hauser1998} and take
their \lq cosmological infrared background', which amounts to half the
observed brightness, the limit comes somewhat close to the curve that is
obtained from WMAP.

The additional curve in the figure (nearly vertical dotted curve) is
the extrapolation of the KBO detected by \citet{Bernstein2004} to
include smaller objects assuming that the population persists to grain
size objects and the power law with the free-parameter 
index specified in the ordinate
for objects smaller than the break radius. The plus symbol on the
curve indicates the power index they inferred for the objects
immediately below the break radius, and the triangle the index inferred
from one object found in the occultation observation
\citep{Schlichting2009}.  The nearly vertical nature of this curve
means that small objects contributes to the total mass only by a
small amount.  It is interesting to note that the anisotropy limit
(with $a_{\rm min}=1$mm)
crosses the extrapolation of the observed objects just
at the canonical slope $\alpha=1.83$ ($q=3.5$).  This means
that the limit on mass from CMB anisotropy experiment is just
consistent with the mass estimate from the observation of KBOs
extrapolated to small size objects for the canonical power index,
although the total mass of observable KBOs is not accurately determined.  
If the small
population of KBO has the size distribution of the canonical power, as
we may see some indication around the Jupiter distance for the \lq
asteroid' grains from the Ulysses and Galileo satellite
\citep{Frisch1999}, emission from KBO is marginally detectable with
CMB anisotropies.  The anisotropy data barely have a power to detect such
a contribution, if not detected with the present data yet.

Finally let us ask what information can be obtained from CMB
anisotropy as to grains in the main asteroid belt. The temperature of
asteroids $\approx 170$K may appear significantly higher than that
relevant to CMB, but these grains should also emit microwaves in the
tail of the Rayleigh-Jeans region.  Following the observation
\citep{Ivezic2001} we take the mass distribution of asteroids that has
the power index $\alpha=2.0$ ($q=4$) for $2.5<a<20$ km and
$\alpha=1.43$ ($q=2.3$) for $0.2<a<2.5$ km with the break radius
$a_{\rm br}=2.5$ km.  We extend the distribution to the Ceres radius
(450km) for large size asteroids in our integration. On the other hand
we extend the distribution to small grains by introducing the possible
second break at the limiting radius of the observation at 0.2 km,
leaving the power index $\alpha$ as a parameter for smaller bodies
beyond the second break radius. The experiment of Ulysses and Galileo
indicates that micrometre-size grains follow the distribution with
$a\approx 1.83$ ($q\approx 3.5$), although no information is available
as to grains larger than a few micrometre size, nor we know if
this grain population is related to the small asteroid population.

\begin{figure}[htbp]
\begin{center}
\includegraphics[scale = 1.5]{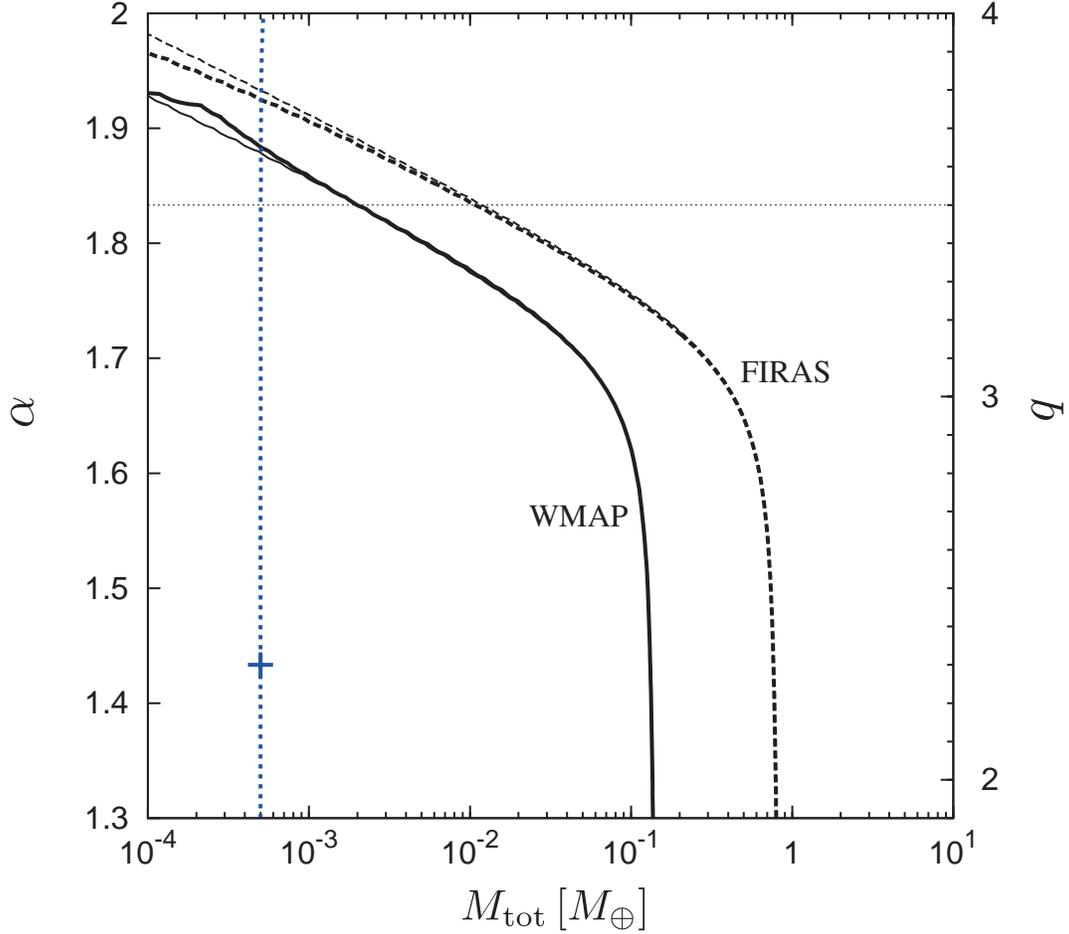}
\caption{Limits on the mass in the asteroid belt as a function
 of the small-mass slope $\alpha$ of the mass function at the 
 95\% C.L. CMB anisotropy constraints from
 WMAP and spectral constraints from
 COBE/FIRAS are shown with solid and dashed curves, respectively.  
 The allowed regions are above the curves.  Thinner lines are
 for $a_{\rm min}=1$\,mm and thicker lines for $a_{\rm
   min}=1$\,$\mu$m.  The nearly vertical dotted line is the total
 mass of the asteroids extrapolated from
 the direct observation with the power index of $\alpha$. The 
 plus symbol is the power index estimated from the observation
 of asteroid just below the (first) break radius. The
 horizontal dotted line refers to the canonical power of collisional
 equilibrium $q=3.5$ of the size distribution.}
\label{fig:MB}
\end{center}
\end{figure}

We carry out a similar calculation for the main asteroid belt at the
average of 2.6\,AU, as did for the microwave emission from KBO, in
order to glance the possible contribution from the main belt.  We
assume the Gaussian distribution with $\sigma=10^\circ$ for the main
asteroid belt, which is broadly consistent with the observation
\citep{Ryan2009}.  We give limits in Figure \ref{fig:MB}, both from
the absence of the spectral distortion constrained by COBE/FIRAS
\citep{Fixsen1996} and of excess anisotropies by WMAP
\citep{Hinshaw2007}, the latter leading to the limit stronger than the
former, again by a factor of 5 in mass. The limit thus derived may be
compared with the estimate of the total mass of asteroids, $\approx
0.0005M_\oplus$ for the observable population
\citep{Asteroids3,Ryan2009}.  It is interesting to note that the limit
on the mass $\lesssim 0.001M_\oplus$ with the canonical power
$q\approx 3.5$ appears close to what is observed for the asteroid
belt.  We remark that the limit with $M_{\rm min}=1$mm (thin curve)
appears below the curve with $M_{\rm min}=1\mu$m (thick curve) in the
small total mass end in the figure due to the cancellation of the
emssivity close to $\alpha\approx 1.93$ arising from the suppression
factor, as we noted above for the case with KBO.

The nearly vertical dotted curve is the  extrapolation of the observed
component of asteroids. This indicates that grains contribute little to
the total mass; it is basically determined by the observed asteroids
($a>0.2$ km).  The contribution to the total mass from
objects smaller than are observable, for instance, 
is only by 1\% if $q=3.5$. The
plus symbol on the curve is the power index derived from the observation of
asteroids smaller than the first break radius down to the observational
limit. The canonical collisional equilibrium power is indicated by the
horizontal dotted line.

The curve of CMB limit shows that $\alpha$ should be smaller than 1.87
($q<3.6$) for the assumed power-law extension of the asteroid population to
smaller sizes, in order to be consistent with the absence of the extra
CMB anisotropy.  The significance of this figure is that the WMAP
limit crosses with the extrapolation from the observed asteroid at
$\alpha$ close to the value that indicates the canonical collisional
equilibrium power or that is inferred from the Ulysses and Galileo
satellite observation.  That is, the limit is just consistent with the
possibility that the grain population detected by those satellites is
the small-size tail of the observed asteroid population.  If the grain
population would follow the power law and continue to what was found by
the Ulysses and Galileo satellite, $q\approx 3.5$ \citep{Frisch1999},
the contribution of the microwave emission is marginally detectable,
if the anisotropy limit could be made tighter by a factor of $2-3$. See
discussion in the next section.

%%%%%%%%%%%%%%%%%%%%%%%%
\section{Discussion and conclusion} \label{sec:conclusion}
%%%%%%%%%%%%%%%%%%%%%%%%

We have shown that the microwave radiation from small bodies in the
Solar System would give excess anisotropy in the microwave sky, and
contribute to the multipole coefficient at low-$\ell$ in excess of the
proper CMB anisotropy.  The current CMB anisotropy experiment, which reaches the
cosmic variance for small $\ell$, already gives an upper limit on the
Edgeworth-Kuiper belt objects stronger than others available to date.
If $q=3.5$, the canonical power index for collisional equilibrium, the
limit is $<0.2\,M_\oplus$, where about 80\% of the mass arises from
objects with $a>15$ km that could be accessible in the optical
observation available to date.  The application to the microwave
emission from the asteroid belt shows that the grain population should
not increase to smaller radii as fast as $q>3.5$. The limit is close to  
the canonical power index. 
If there is a population that interpolates asteroids and
grains found by Ulysses-Galileo satellite, it would cause a signal
marginally detectable when the anisotropy limit would be improved,
say, by a factor of 2. In summary, we would underline that the current
limit set from CMB anisotropy is just above what could be significant
to understand the world of small asteroids and grains for both
Edgeworth-Kuiper belt and main asteroid belt.  Tightening of the anisotropy
limit by a factor of 2 or 3 would bring us a significant new insight.

\begin{figure}[!t]
\begin{center}
\includegraphics[scale = 1.1]{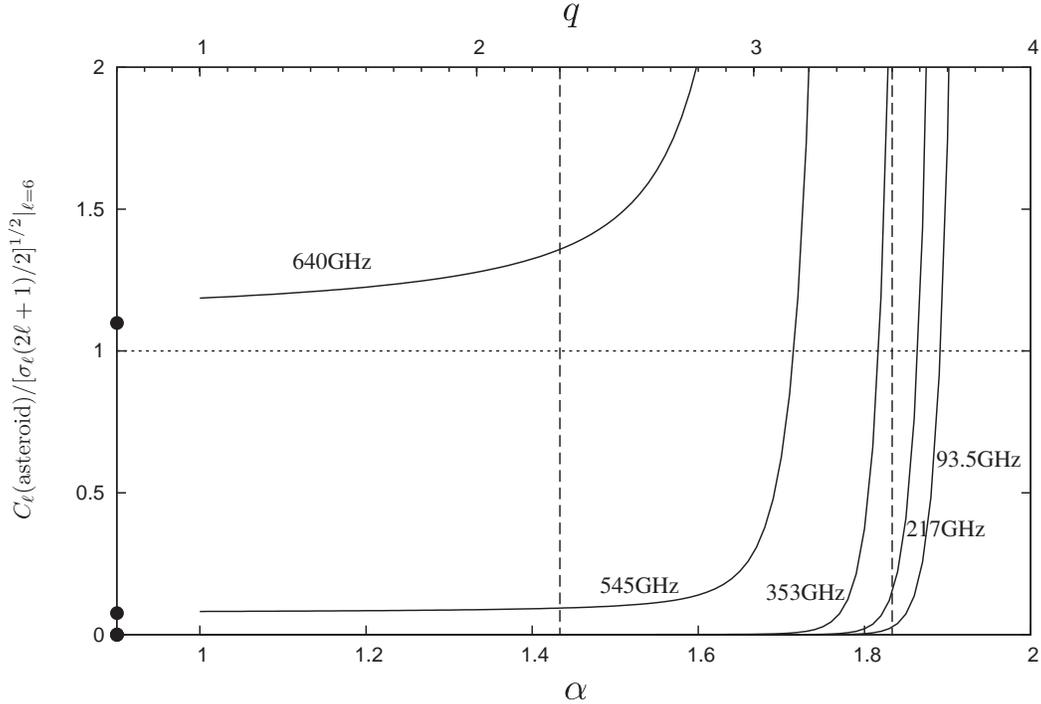}
\caption{Expected contribution of the 
asteroid signal to the CMB multipole
compared with the cosmic
variance $C_\ell({\rm asteroid})/[\sigma_\ell(2\ell+1)/2]^{1/2}$ at 
the multipole, $\ell=6$, as a function of the power index of the
small-size objects $\alpha$ for several choices of frequency $\nu$. 
The symbols on the
left ordinate indicate the expected asteroid signal from the observed
population ($a>0.2$km) of asteroid belt alone. 
The dotted line is the canonical power of collisional
equilibrium $q=3.5$ (or the power index of grains detected in the
Ulysses Galileo mission) and the power of smaller observed asteroids.}
\label{fig:ell6MB}
\end{center}
\end{figure}

The current limit from microwave emission from asteroids is derived
taking the cosmic variance, which was already reached by WMAP for
$\ell$ that concerns us.  This means that we are not readily able to strengthen
the limit by improving the experimental accuracy.  We may think
of two ways, however, to improve the limit or possibly detect the
signal. One is to work with higher frequency (e.g., in the Planck
satellite, which observes, e.g., at 217, 353, 545 and 857 GHz) 
where the relative importance of grain emission increases
with $\nu^2$, as anticipated in the Rayleigh-Jeans region, while the
CMB is in the Wien region beyond 150 GHz.  Repeating similar
calculations, we have confirmed that the CMB anisotropy limit goes
beyond what is expected from the extrapolation of the KBO observations
with the canonical power of $q=3.5$ if the 353 GHz observation with
the Planck satellite \citep{Lamarre2010} would not see any signals
beyond the cosmic variance. 

Similarly, the CMB anisotropy limit on the asteroids in the main belt
also comes beyond the crossing point of the extrapolation of observed
asteroid with the $q=3.5$ power if 353 GHz is used.  Figure
\ref{fig:ell6MB} shows the asteroid signal compared with the cosmic
variance, $C_\ell({\rm asteroid})/[\sigma_\ell(2\ell+1)/2]^{1/2}$, at
an optimal multipole for the main asteroid belt, $\ell=6$, as a
function of the power index of the small-size objects $\alpha$ for
several choices of frequency $\nu$ (we took some of them as values
used by the Planck mission). The symbols on the left ordinate are the
expected CMB signals from the observed population of asteroids
alone. The curves include the contributions from small objects
extrapolated with a power law with the index $\alpha$ down to 1 mm
size. This shows that the signal appears beyond the cosmic variance
for the observation with $\nu\gtrsim 350$ GHz if small size objects
obey the power law with the canonical power: for $ \nu\gtrsim 640$ GHz
the emission from the observed population of asteroids alone will
exceed the cosmic variance at the $\ell=6$ multipole.  This brief
analysis would warrant more realistic modelling of the microwave
emission from the main asteroid belt, including its three dimensional
structure, and a more detailed analysis.

The other way to tighten the limit may be to use a templet filter that
matches the spatial distribution of objects in the Edgeworth-Kuiper belt or
the asteroid belt to enhance their signals. A possibility is not yet
excluded for a detection of microwave emission from small bodies
in anisotropy measurements.

\begin{acknowledgements}
We would like to thank Bruce Draine for useful comments.
\end{acknowledgements}

%%%%%%%%%%%%%%%%%%%%%%%

\end{document}